# High-Efficiency Ising Machine with Time-Dimensional Exchange Coupling

Zhengyu Du[1#], Haijie Xu[2#], Kaiming Cai[3], Zhe Yuan[2†], Yue Zhang[1††]

1. School of Integrated Circuit, Huazhong University of Science and Technology, Wuhan, China.
2. State Key Laboratory of Surface Physics and Interdisciplinary Center For Theoretical Physics and Information Sciences, Fudan University, Shanghai, China.
3. School of Physics, Huazhong University of Science and Technology, Wuhan, China.

# These authors contributed equally to this work.

† Contact author: yuanz@fudan.edu.cn (Zhe Yuan)

†† Contact author: yue-zhang@hust.edu.cn (Yue Zhang)

Conventional probabilistic Ising machines often suffer from inefficient exploration of configuration space, while replica-based quantum Monte Carlo methods reduce sampling bottlenecks at the cost of large hardware overhead. Here we propose a time-dimensional exchange coupling (TEC) that replaces spatial replica coupling with a temporal exchange interaction between successive spin configurations of a single p-bit network. This TEC improves sampling efficiency without duplicating replica hardware. At low temperatures, antiferromagnetic TEC expands sampling range; at high temperatures, ferromagnetic TEC stabilizes the optimal state. For MaxCut problems with up to 2000 vertices, TEC significantly accelerates convergence speed. SPICE simulations confirm hardware feasibility. TEC offers a scalable, hardware-efficient strategy to enhance combinatorial optimization on existing Ising machines.

The Ising model, originally introduced to describe magnetic phase transitions, has become a paradigmatic framework for combinatorial optimization problems. In this mapping, the core problem is to determine the ground state of an Ising Hamiltonian $H = \sum_{i<j} J_{ij}\sigma_i\sigma_j$ . Here $J_{ij}$ indicates the exchange coupling between two spins $\sigma_i$ and $\sigma_j$ with different spatial coordinates. The ground state of the Ising Hamiltonian encodes the correct solution to problems such as MaxCut, quadratic unconstrained binary optimization, and many NP-hard tasks [1].

Building on the Ising model, several probabilistic computing approaches have been developed to tackle these combinatorial optimization problems. These include simulated annealing (SA), simulated quantum annealing, parallel tempering, and algorithms based on probabilistic bits (p-bits)—fluctuating two-state units that emulate thermal spins via tunable random telegraph noise [2-8]. The p-bit algorithms increase the probability for escaping the local minimum trap, which favors the chance for finding the global optimal solution.

However, a common challenge across these p-bit algorithms is that the searching efficiency critically depends on how well the sampling range over spin configurations is modulated. For p-bit-based sampling, such as the Monte Carlo method, adjusting the effective noise temperature or the energy barrier in the search can significantly influence the probability to reach the ground state. Therefore, improving computational efficiency crucially lies in the adaptive control of the sampling scope—balancing exploration of the vast configuration space against exploitation of promising low-energy regions. Nevertheless, in conventional p-bit algorithms, introducing strong randomness to escape the local minimum trap is usually accompanied with the weakening of the stability of the global optimal state.

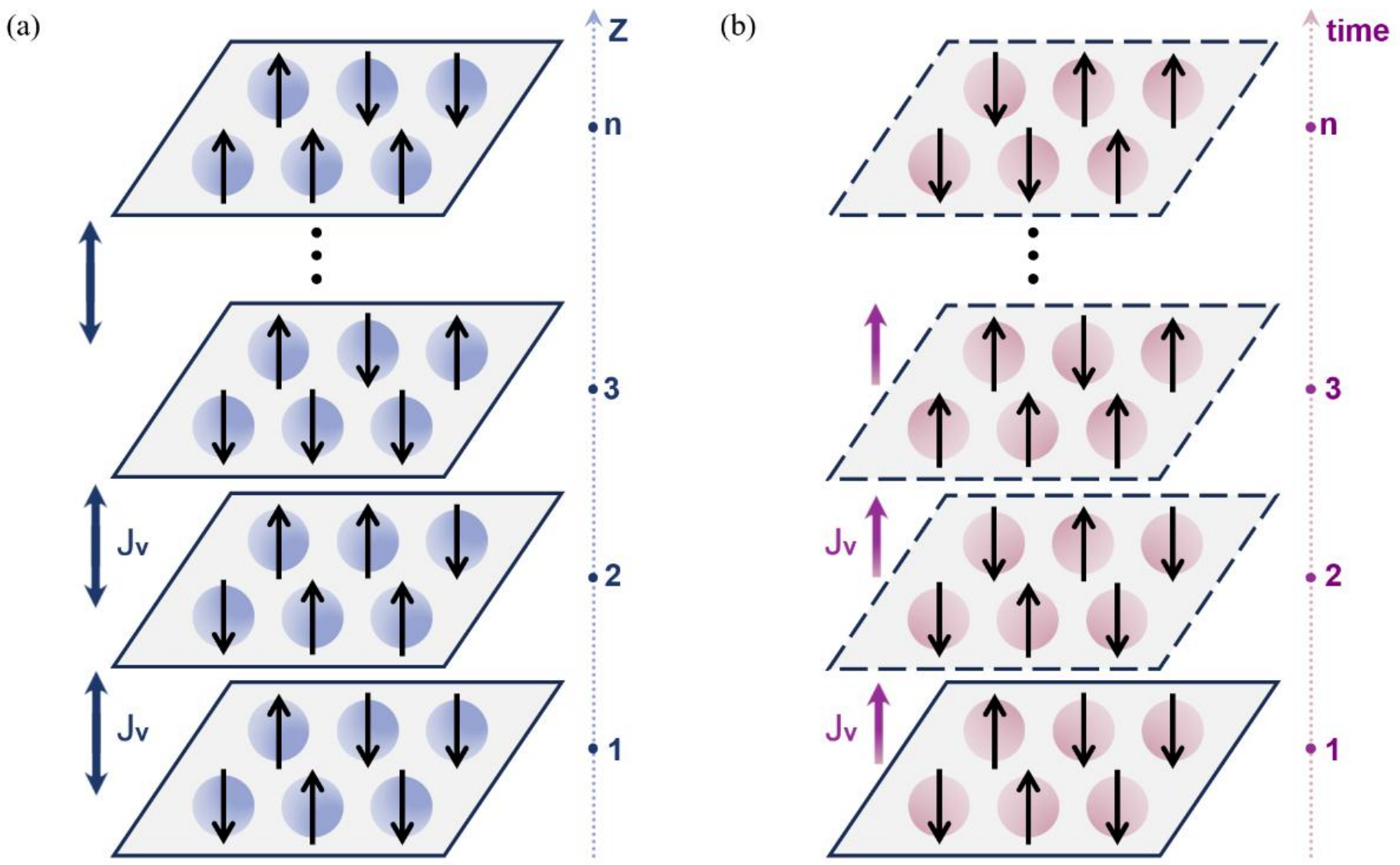

**FIG. 1. (a) Conventional spatial coupling in QMC with symmetric, bidirectional interlayer coupling $J_v$ along the auxiliary $z$ axis. (b) The proposed Time-Dimensional Exchange Coupling (TEC), featuring unidirectional, causal feedback $J_v$ from the prior state along the time axis.**

Quantum Monte Carlo (QMC) methods, originating from quantum many-body physics, have emerged as typical algorithms for tackling such low-efficiency exploration in p-bit-based sampling [9]. A representative example of QMC is the multi-replica coupling model with inter-replica exchange (parallel tempering), which maps a quantum many-body Hamiltonian onto a classical p-bit counterpart [11-13], as illustrated in Fig. 1(a). By simulating multiple replicas of the system at different temperatures and allowing configuration exchanges, QMC enhances ergodicity and helps escape metastable states—addressing the challenge of complex spin–spin coupling landscapes. However, when implementing such "coupled-replica" Ising machines in hardware, each replica requires its own physical spin array. This leads to a multiplicative increase in hardware footprint, posing a severe constraint for large-scale integration, and reintroduces the spatial coupling bottlenecks that several p-bit hardware architectures were designed to overcome [14-16].

Inspired by the spatial replica coupling, we propose a fundamentally different strategy: replace the spatial coupling among multiple parallel replicas with a time-dimensional exchange coupling (TEC) between successive sampling instances of an Ising machine with a single p-bit network [Fig. 1(b)]. Specifically, we replace the spatial inter-replica exchange coupling by an exchange-like interaction term that couples the spin configuration at the $(t-1)^{\text{th}}$ sampling time step to that at the $t^{\text{th}}$ step, forming a one-dimensional chain along the time axis. This TEC correlates the flip probability of p-bit with the history of the spin configuration, biasing the Markov chain toward collective moves that explore a broader region of configuration space while stabilizing the correct ground state.

We demonstrate the effectiveness of this approach on benchmark MaxCut problems for up to 2000 vertices [17]. Using a simulated Ising machine with TEC, we observe a substantial improvement in both the probability of finding the exact solution and the convergence rate. Our results suggest that TEC offers a scalable, hardware-friendly alternative to replica-based methods, promising a new class of Ising machines that harness temporal correlations rather than spatial parallelism. These findings may open the door to compact and efficient solvers for large-scale combinatorial optimization.

In the following, we briefly describe the theoretical framework and outcomes for introducing the TEC. We consider a system of $N$ binary spins $\sigma_i(t)$ = 1, −1 evolving in discrete time steps. The instantaneous spatial energy is given by

$$E_{\text{spatial}}[\boldsymbol{\sigma}(t)] = -\sum_{i<j} J_{ij}\sigma_i(t)\sigma_j(t)\,. \tag{1}$$

To introduce TEC, we define the total energy by adding an exchange coupling term between consecutive configurations:

$$E_{\text{total}}[\boldsymbol{\sigma}(t),\boldsymbol{\sigma}(t-1)] = E_{\text{spatial}}[\boldsymbol{\sigma}(t)] - J_v\sum_i \sigma_i(t)\sigma_i(t-1), \tag{2}$$

where $J_v$ is the strength of the TEC. Here the iteration between the $t-1$ and $t$ step is assumed to be much longer than that for the spin relaxion.

We extend the Glauber-type stochastic dynamics to account for spin updates influenced by TEC [18, 19]. The TEC is found to modify the local update velocity of the magnetization $q_i(t)$ at site $i$ as

$$\left[\frac{dq_i(t)}{d\alpha t}\right]_{J_v} = \left(\frac{1-\eta^2}{1+\eta^2}\right)\tanh\left[\frac{J_v\sigma_i(t-1)}{k_B T}\right] \tag{3}$$

Here $\alpha$, $k_B$, and $T$ are the switching rate, Boltzmann constant, and absolute temperature, respectively. $\eta = \tanh\left(\frac{2J}{k_B T}\right)$ with $J$ the exchange energy between the neighboring spins. More details about the derivation of Eq. (3) are in the Supplementary Materials. Eq. (3) indicates TEC either accelerates or decelerates spin update, depending on the sign of $J_v$ and the spin state in the preceding time step.

The velocity modulation directly affects the spin-flip probability at thermal equilibrium. Under the Boltzmann–Maxwell distribution, the flip probability of spin $i$ at time $t$ takes the form

$$P_i^{\text{flip}}(t) = \frac{1}{1+\exp\{2\beta\sigma_i(t)[h_i^s(t)+h_i^t(t-1)]\}}, \tag{4}$$

where $h_i^{\text{spatial}} = \sum_j J_{ij}\sigma_j(t)$ is the spatial exchange field from the spins at other sites, and $h_i^t = J_v \sigma_i(t-1)$ is the temporal exchange field from the previous state of spin *i*.

In the low-temperature limit (large $\beta$), the large energy barrier simplifies the flip probability to:

$$P_i^{flip}(t) \approx R_{TEC} \exp[-2\beta\sigma_i(t)h_i^s(t)], \tag{5}$$

where $R_{TEC} = \exp[-2\beta J_v \sigma_i(t)\sigma_i(t-1)]$, demonstrating that TEC introduces a self-modulated bias proportional to the previous spin state and exponentially modifies the flip probability.

The temporal spin configuration exhibits ferromagnetic (FM) or antiferromagnetic (AFM) ordering depending on the sign of $J_v$. For $J_v < 0$ (AFM TEC), consecutive spin states tend to be antiparallel, promoting frequent flipping at each subsequent step and thereby substantially expanding the exploration range in configuration space. Conversely, for $J_v > 0$ (FM TEC), each spin tends to retain its previous state, suppressing updates. Perfect temporal FM or AFM order is realized at $T = 0$ K. At finite temperature, stochasticity enters through the spin-flip probability [Eq. (4)].

In conventional Ising machines without TEC, the spin configuration often becomes trapped in incorrect local minima at low temperatures or escapes from the global minimum due to strong thermal fluctuations at high temperatures. The TEC-based Ising model addresses this issue by effectively modulating the spin-update rate. Introducing an AFM TEC enhances the probability of escaping local minima at low temperatures, whereas an FM TEC stabilizes the global minimum against thermal fluctuations at high temperatures.

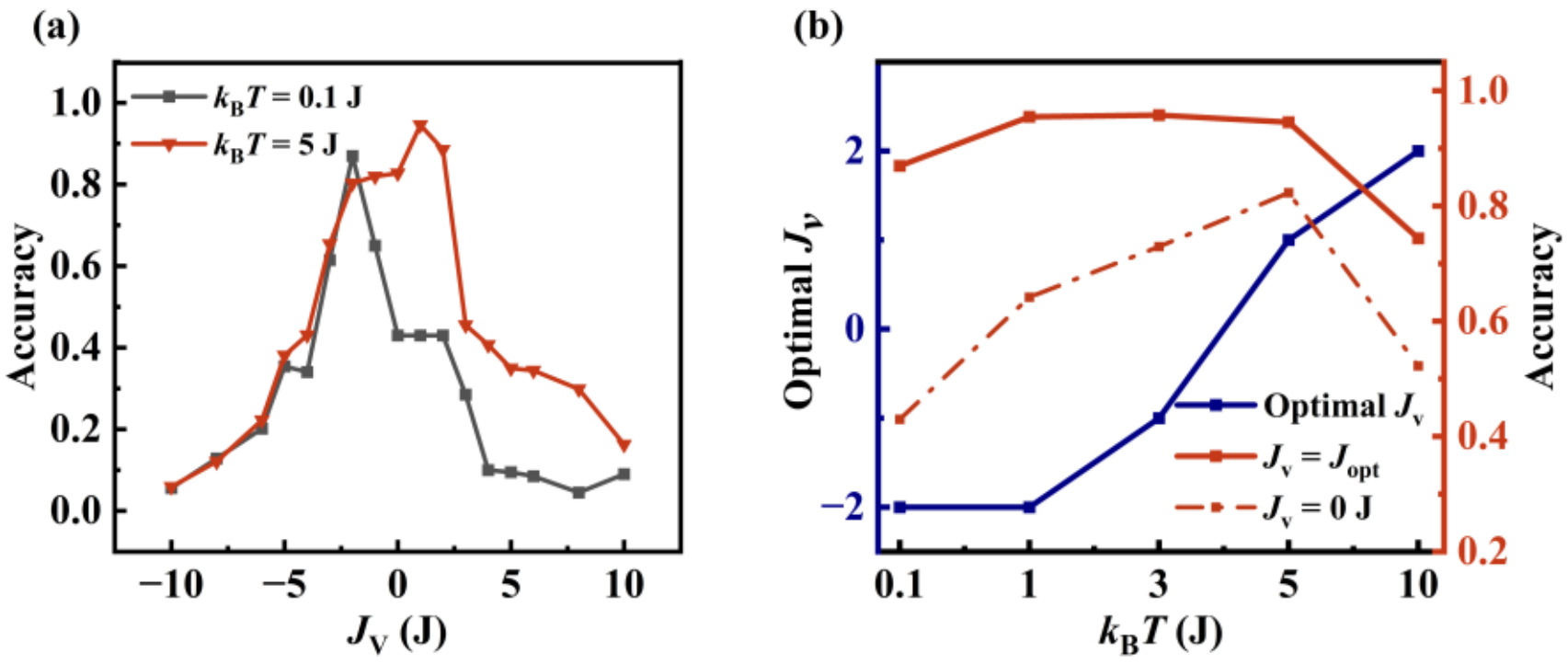

**FIG. 2. Performance of the TEC in solving a 10-node Max-Cut instance problem. (a) Solution accuracy versus TEC coupling $J_v$ at different effective temperatures ($k_B T = 0.1$ and 5 J). (b) Evolution of the optimal coupling [the $J_v$ for the maximum accuracy in (a)] and its corresponding peak accuracy as a function of $T$ (The red dashed line corresponds to the peak accuracy without TEC.).**

We validate this improvement through numerical simulations of a MaxCut problem, with results summarized in Fig. 2. For a representative 10-node instance, we plot the solution accuracy as a function of the TEC coupling strength $J_v$ at two distinct temperatures [Fig. 2(a)]. At low simulated temperature ($k_BT$ = 0.1 J), the optimal $J_v$ is negative, while at high temperature ($k_BT$ = 5 J), it becomes positive. This trend is systematically corroborated in Fig. 2(b), where the optimal $J_v$ shifts monotonically from negative to positive with increasing temperature—consistent with the theoretical picture that negative $J_v$ broadens the exploration range to facilitate escape from local minima in the low-temperature regime, whereas positive $J_v$ helps preserve the correct state when thermal agitation is already substantial. Additionally, Fig. 2(b) compares the accuracy with optimal $J_v$ and without TEC, demonstrating that the TEC cases outperform that without TEC across the entire temperature range.

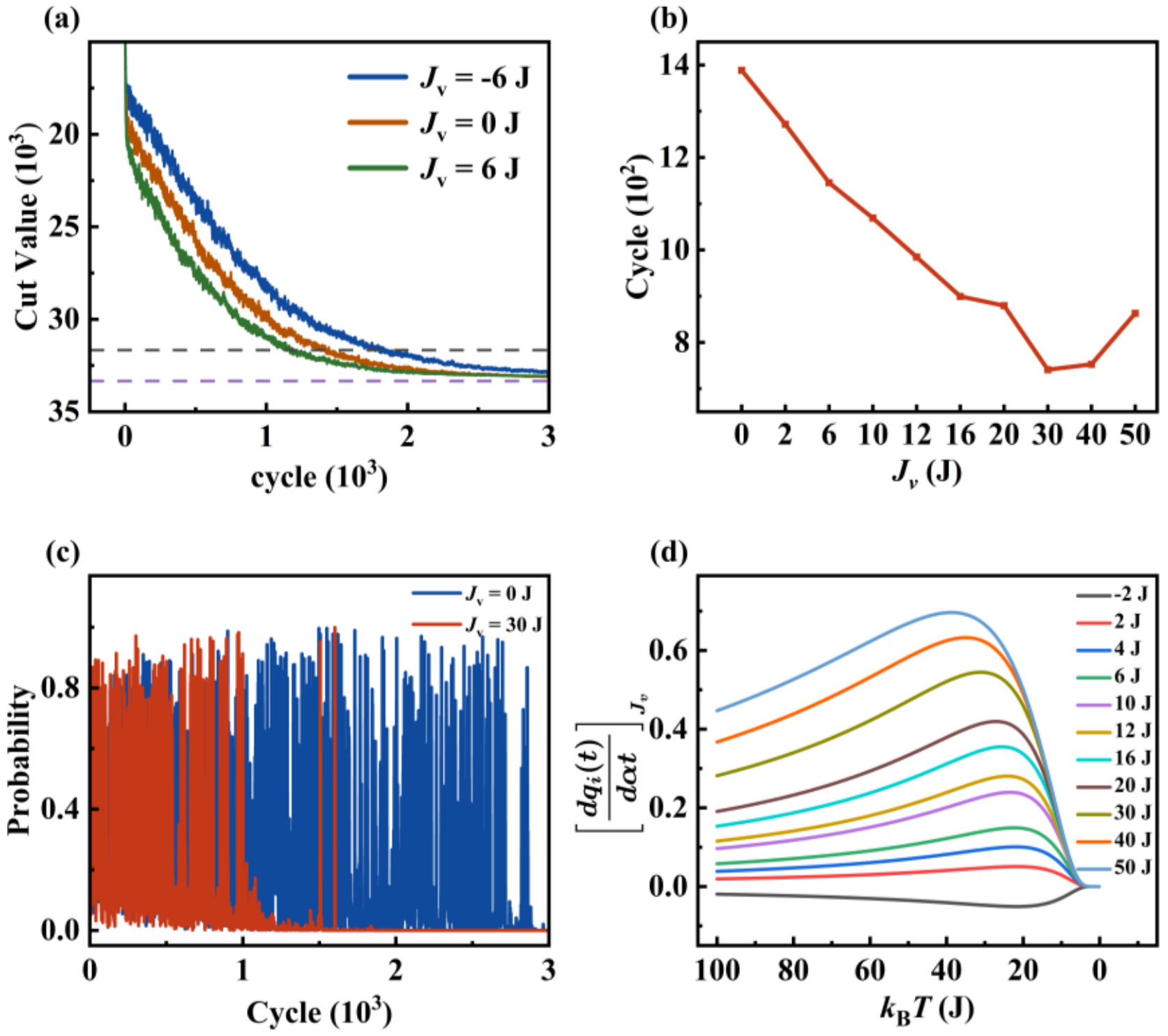

**FIG. 3. Annealing dynamics and convergence acceleration for a large-scale K2000 benchmark instance. (a) Evolution of the cut value over annealing cycles under various temporal coupling strengths $J_v$. The inset provides a magnified view of the early-stage trajectories, demonstrating the acceleration of the initial descent towards the ground state. (b) The number of cycles required to reach 95% of the optimal cut value (31670) as a function of $J_v$. (c) The evolution of flip probability with cycles for $J_v$ = 0 and 30 J based on the calculation of Eq. (4). (d) The TEC-induced update velocity with temperature at different $J_v$ based on the calculation of Eq. (3).**

To fully demonstrate the improvement afforded by the TEC in solving the MaxCut problem, we present benchmark results for a 2000-node instance, obtained using simulated annealing with TEC (Fig. 3) [17]. Figure 3(a) displays the annealing trajectories for $J_v$ = 0 and $J_v$ = ± 6 J, over a temperature range from $k_BT$ = 100 J down to 0.1 J. For all considered $J_v$, every run converges to the exact solution [cut value $1.67 \times 10^3$, the purple dashed line in Fig. 3(a)] with a success rate exceeding 95% [the black dashed line in Fig. 3(a)]. However, relative to the $J_v$ = 0 case, positive (negative) $J_v$ accelerates (decelerates) the convergence. Figure 3(b) shows the number of annealing cycles required to reach 95% accuracy as a function of $J_v$. The optimal value, $J_v$ = 30 J, reduces the convergence time by approximately half compared to the case without TEC. This acceleration is corroborated by calculations of the iterative flip probability given in Eq. (4) [Fig. 3(c)]. These results further confirm the effectiveness of TEC in enhancing both the quality and efficiency of Ising-based optimization.

The nonmonotonic behavior observed in Fig. 3(b) arises from the intricate temperature dependence of the update velocity in Eq. (3). The velocity vanishes at both zero and infinite temperatures, with a maximum expected at an intermediate temperature. Fig. 3(d) presents calculated update velocities under different TEC parameters, assuming the spin state at the $(t-1)^{\mathrm{th}}$ step is fixed to 1. As anticipated, a negative (positive) $J_v$ reduces (increases) the update velocity, and the temperature at which the maximum occurs shifts to lower values with increasing $J_v$.

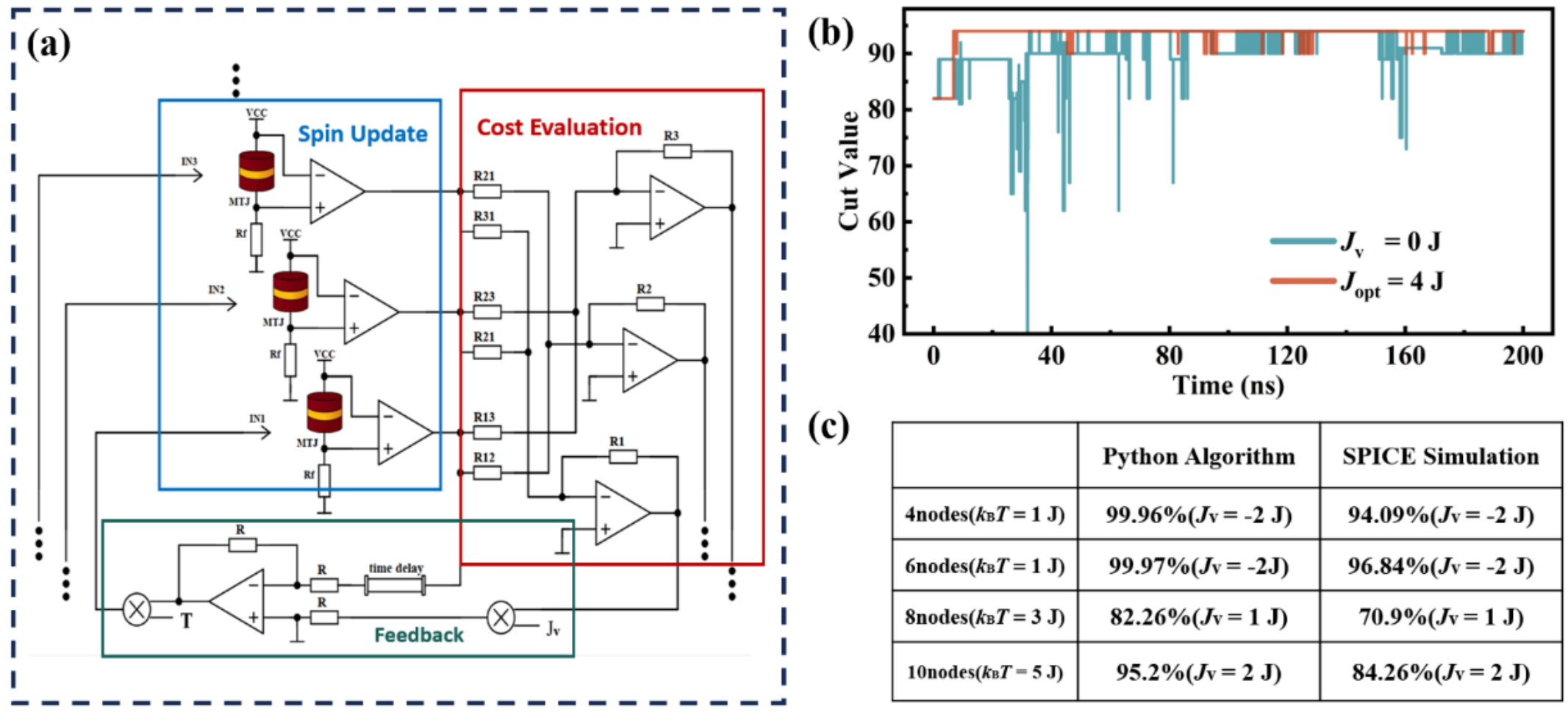

| | Python Algorithm | SPICE Simulation |
|---|---|---|
| 4nodes($k_BT$ = 1 J) | 99.96%($J_v$ = -2 J) | 94.09%($J_v$ = -2 J) |
| 6nodes($k_BT$ = 1 J) | 99.97%($J_v$ = -2J) | 96.84%($J_v$ = -2 J) |
| 8nodes($k_BT$ = 3 J) | 82.26%($J_v$ = 1 J) | 70.9%($J_v$ = 1 J) |
| 10nodes($k_BT$ = 5 J) | 95.2%($J_v$ = 2 J) | 84.26%($J_v$ = 2 J) |

**FIG. 4. (a) Schematic of the closed-loop hardware solver, comprising stochastic magnetic tunnel junctions (s-MTJs) for spin sampling, a resistive crossbar for Hamiltonian evaluation, and time-delayed feedback ($J_v$) to active perturb local energy minima. (b) Transient dynamics for an 8-node Max-Cut instance. The uncoupled network ($J_v$ = 0 J) exhibits persistent stochastic wandering without stabilization, whereas the optimal feedback ($J_{opt}$ = 4 J) guides the system to sustain the ground state. (c) Success probabilities across varying graph scales (4 to 10 nodes). The acceptable accuracy degradation in SPICE simulations compared to the ideal Python algorithm (e.g., 70.9% vs. 82.26% for 8 nodes) originates from intrinsic circuit non-idealities, confirming the physical robustness of the framework.**

To verify the feasibility for carrying out the TEC in a hardware, we did hardware simulation by using Simulation Program with Integrated Circuit Emphasis (SPICE) (Fig. 4). The proposed physical solver is a closed-loop dynamical system that synergizes intrinsic stochastic fluctuations with massively parallel physical interactions. The core of each Ising node is a magnetic tunnel junction (MTJ)-based p-bit, which serves as the physical entropy source [13]. By biasing the MTJ near its instability region, thermal fluctuations are transcoded into random telegraph noise. A comparator is then used to threshold this analog signal, yielding a fluctuating binary output.

The spatial interactions between spins are encoded into a dense analog resistive network [Fig. 4(a)], which serves as a direct physical mapping of the problem's adjacency matrix $G_{ij}$. Through this analog architecture, a massive parallel weighted summation of the spin states is executed to continuously evaluate the local field $h_i = \sum J_{ij}\sigma_j$ and the total energy (cut value) of the current configuration. By emulating the system's Hamiltonian, this network provides an instantaneous readout of the underlying energy landscape, directly supplying the essential deterministic driving force required for the subsequent adaptive state evolution.

To prevent the system from becoming trapped in local energy minima, an adaptive feedback loop usually modulates the spin update process [20]. In this work, the feedback circuit serves as the direct physical realization of the TEC $J_v\sigma_i(t-1)$. By incorporating a hardware time delay, the previous spin state is preserved and fed back into the node as an effective temporal bias. Scaled by the TEC strength $J_v$ and the effective thermal energy $k_BT$, this delayed feedback actively perturbs configurations, effectively steering the collective dynamics of the network toward the global optimal Max-Cut solution.

The physical efficacy and scalability of this architecture are validated through transient simulations and algorithmic benchmarking. As dynamically visualized in an 8-node instance [Fig. 4(b)], the uncoupled network ($J_v = 0$ J) suffers from persistent stochastic wandering; although it temporarily visits the global optimum, it fails to stabilize. Conversely, introducing the optimal TEC ($J_{opt} = 4$ J) provides a crucial stabilizing mechanism. This delayed feedback effectively counteracts excessive stochastic fluctuations, significantly enhancing the system's ability to locate and sustain the ground state.

Quantitative cross-validation across 4-to-10-node graphs [Fig. 4(c)] reveals a highly consistent scaling trend between SPICE simulations and the Python algorithm. The minor accuracy degradation in SPICE stems from intrinsic circuit non-idealities, such as finite noise limits and analog component variations. Ultimately, this acceptable discrepancy highlights the framework's robustness, proving that high-fidelity combinatorial optimization can be achieved using hardware components.

In summary, this work introduces a TEC for probabilistic Ising machines, addressing the inefficiency of conventional p-bit sampling and the hardware overhead of spatial replica methods like parallel tempering. By coupling successive spin configurations along the time axis, TEC enables controlled exploration versus exploitation. AFM TEC enhances escape from local minima at low temperatures, while FM TEC stabilizes solutions at high temperatures. For MaxCut benchmarks, numerical results show > 95% success rates and accelerated annealing schedules. Hardware simulations using MTJ-based p-bits and SPICE confirm physical realizability. The work provides a scalable, hardware-friendly alternative to replica-based Ising machines.

**Acknowledgements**
This work was supported by the National Natural Science Foundation of China (Grant No. 12574119), the open research fund of Songshan Lake Materials Laboratory (Grant No. 2023SLABFN26), and the Wuhan City Key R&D Program (Grant No. 2025050602030069).